\documentclass[12pt]{article}
\usepackage{bbold}
\usepackage{slashed}
\usepackage{tikz}
\usepackage{graphicx}
\usepackage{epstopdf}
\usepackage{subfigure}
\usepackage{amsmath}
\usepackage{amsfonts}
\usepackage{amssymb}
\usepackage{color}
\usepackage{mathrsfs}
\usepackage[colorlinks=true, linkcolor=blue, citecolor=blue, urlcolor=blue]{hyperref}

\numberwithin{equation}{section}

\newcommand{\beq}{\begin{equation}}
\newcommand{\eeq}{\end{equation}}
 
\newcommand{\bea}{\begin{eqnarray}}
\newcommand{\ea}{\end{eqnarray}}
\newcommand{\barr}{\begin{array}}
\newcommand{\earr}{\end{array}}
\newcommand{\lb}{{\langle}}
\newcommand{\rb}{{\rangle}}

\newcommand{\calo}{{\cal O}}

\newcommand{\zb}{\overline{z}}

\begin{document}
\begin{titlepage}

\setcounter{page}{1} \baselineskip=15.5pt \thispagestyle{empty}

%\begin{flushright}
%hep-th/13mmnnn\\
%\end{flushright}
\vfil
\begin{center}

\def\thefootnote{\fnsymbol{footnote}}
\begin{changemargin}{0.05cm}{0.05cm} 
\begin{center}
{\large \bf %Lyapunov from Goldstone:\\[4mm] 
\LARGE{On CFT and Quantum Chaos}}
\end{center} 
\end{changemargin}

~\\[1.5cm]
{ Gustavo J. Turiaci${}^{\rm a}$ and Herman L. Verlinde${}^{\rm a,b}$}\footnote{e-mail:  turiaci@princeton.edu, verlinde@princeton.edu}
\\[0.3cm]

{\normalsize { \sl ${}^{\rm a}$Physics Department and ${}^{\rm b}$Princeton Center for Theoretical Science %Princeton University Princeton, NJ 08544, USA
\\[3mm]
Princeton University, Princeton, NJ 08544, USA}}

\end{center}

\vspace{1cm}

%\hrule
 \vspace{1cm}
\begin{changemargin}{01cm}{1cm} 
{\small  \noindent 
\begin{center} 
\textbf{Abstract}
\end{center} }
We make three observations that help clarify the relation between CFT and quantum chaos. We show that any 1+1-D system in which conformal symmetry is non-linearly realized exhibits two main characteristics of chaos:  maximal Lyapunov behavior and a spectrum of Ruelle resonances. We use this insight to identify a lattice model for quantum chaos, built from parafermionic spin variables with an equation of motion given by a Y-system. Finally we point to a relation between the spectrum of Ruelle resonances of a  CFT and the analytic properties of OPE coefficients between light and heavy operators. In our model, this spectrum agrees with the quasi-normal modes of the BTZ black hole.
\end{changemargin}
 \vspace{0.3cm}
%\hrule
\vfil
\begin{flushleft}
\today
%March 20, 2013
\end{flushleft}

\end{titlepage}

%\newpage
%\tableofcontents
%\newpage

\addtolength{\abovedisplayskip}{.5mm}
\addtolength{\belowdisplayskip}{.5mm}

\def\plus{\raisebox{.5pt}{\tiny$+$\smpc}}

\addtolength{\parskip}{.6mm}
\def\spc{\hspace{1pt}}

\def\nspc{{\hspace{-1pt}}}
\def\ff{\rm\smpc f\smpc} 
\def\fff{\mbox{Y}}
\def\ww{{\rm w}}
\def\smpc{{\hspace{.5pt}}}

\def\zz{{\spc \rm z}}
\def\xx{{\rm x\smpc}}
\def\xxi{\mbox{\footnotesize \spc $\xi$}}
\def\jj{{\rm j}}
 \addtolength{\baselineskip}{-.1mm}

\renewcommand{\Large}{\large}

\section{Introduction}\label{sec:intro}
There has been considerable recent interest in the manifestation of many body quantum chaos in strongly coupled conformal field theory \cite{shenker-stanford, JMV, maximal}.
Characteristics of chaotic systems, such as Lyapunov behavior, scrambling and  Ruelle resonances, can be effectively isolated by studying out-of-time ordered correlation functions \cite{SYK, Polchinski:2015cea, morechaos}. AdS/CFT duality relates these characteristics to evident properties of wave perturbations near black hole horizons, such as exponential redshifts, gravitational shockwaves and quasi-normal modes.

Many body quantum chaos is interesting in its own right, but usually hard to~quantify. Identifying simple models or general mechanisms that exhibit aspects of quantum chaos is therefore a worthwhile goal. In this note we make three interrelated observations that may help 1) identify a new class of toy models in the form of a simple lattice model built out of parafermionic spin variables
2) clarify the relationship between maximal quantum chaos and the non-linear realization of conformal symmetry at finite temperature,
3) relate the spectrum of Ruelle resonances to analytic properties of OPE coefficients in the CFT. We now briefly describe each of the three components of our story.
\medskip
\medskip
\def\is{\! & \! = \! & \!}
\def\CC{{\rm C}}

 {\it 1) A discrete model of many body quantum chaos} 
 
 Useful many body systems that may exhibit chaos are quantum spin chains and matrix models.  Another interesting example is the SYK %Sachdev-Ye-Kitaev 
 model, which is solvable at strong coupling, maximally chaotic, and exhibits emergent conformal symmetry at low energies \cite{SYK}. Our model of interest combines ingredients and properties of both examples, with the added feature that its Lyapunov behavior can be exhibited via weakly coupled effective field theory.  The model described below is a minor specialization of the class of integrable lattice models introduced by Faddeev, Kashaev and Volkov \cite{Faddeev, FT, Teschner-LL}. %}

\def\bea{\begin{eqnarray}}

\def\eea{\end{eqnarray}}
\def\NN{\mbox{\footnotesize\nspc\smpc\sc n}}
\def\WW{\mbox{\footnotesize\nspc\sc w}}
\def\SS{\mbox{\footnotesize\nspc\smpc\sc s}}
\def\EE{\mbox{\footnotesize\nspc\smpc\sc e}}
\def\YY{\mbox{\footnotesize \sc y}}
\def\eps{{\mbox{\footnotesize \nspc$\epsilon$}}}
\def\UU{{\rm U}}
\def\VV{{\rm V}}
The model is assembled from a collection of $\mathbb{Z}_N$ parafermionic operators $\ff_n$, labeled by an integer $1 \leq n \leq L$ with $L$ some large odd integer. We identify $\ff_{L+1}\equiv \ff_1$, so the integers n label points on a 1D periodic lattice. The $\ff_n$ satisfy the algebra
\beq
\label{parafermion}
\qquad \ff_{2n\pm 1} \ff_{2n}\spc  = \spc q^2\, \ff_{2n} \spc \ff_{2n\pm 1}, \qquad \qquad
q = e^{{i \pi}/{N}},
\eeq
while $[\ff_n,\ff_m] = 0$ for $|m-n|\geq 2.$ This parafermion algebra can be realized on a finite dimensional Hilbert space ${\cal H} = V_1 \otimes V_2 \otimes...\otimes V_{L}$ with $V_n$ an $N$-dimensional vector space attached to the link between site $n$ and $n+1$, on which $\ff_n$ and $\ff_{n+1}$ act via appropriate clock and shift matrices. In the end, we imagine taking the continuum limit $L\to \infty$.  The integer $N$  is assumed to be large but finite.\footnote{As we will see shortly, $N$ will be proportional to the central charge of the  low energy effective CFT.}

\begin{figure}[t!]
  \begin{center}
  \begin{tikzpicture}[scale=0.8]
    \coordinate (Origin)   at (0,0);
    \coordinate (XAxisMin) at (-2.5,-2.25);
    \coordinate (XAxisMax) at (-0.5,-2.25);
    \coordinate (YAxisMin) at (-2.25,-2.5);
    \coordinate (YAxisMax) at (-2.25,-0.5);
    \draw [line width=0.75mm, black,-latex] (XAxisMin) -- (XAxisMax);
    \node[] at (-0.5,-2.6) {$\sigma$};
   % \node[pos=1, above]{$t$};
    \draw [line width=0.75mm, black,-latex] (YAxisMin) -- (YAxisMax);
    \node[] at (-2.6,-0.5) {$\tau$};
    \clip (-3.3,-3.3) rectangle (2.5cm,2.5cm); 
    
% Puts the shaded rectangle
\foreach \x in {-3,-2,...,3}{
\foreach \y in {-3,-2,...,3}{
\node[draw,circle,inner sep=2.5pt,fill] at (1.5*\x,1.5*\y) {};}} % Places a dot at those points 
%Edges 
\node[draw,circle,inner sep=3pt,fill, blue] at (-1.5,0) {};
\node[] at (-2.1,0) {$Y_{\rm W}$};
\node[draw,circle,inner sep=3pt,fill, blue] at (0,-1.5) {};
\node[] at (0,-2.1) {$Y_{\rm S}$};
\node[draw,circle,inner sep=3pt,fill, blue] at (0,1.5) {};
\node[] at (0,2.1) {$Y_{\rm N}$};
\node[draw,circle,inner sep=3pt,fill, blue] at (1.5,0) {};
\node[] at (2.1,0) {$Y_{\rm E}$};
%Center
\node[] at (0,0.5) {$\small{(\sigma,\tau)}$};
%Diamons 
\path[draw, line width=2pt,blue] (-1.5,0) -- (0,-1.5);
\path[draw, line width=2pt,blue] (0,-1.5) -- (1.5,0);
\path[draw, line width=2pt,blue] (1.5,0) -- (0, 1.5);  
\path[draw, line width=2pt,blue] (0,1.5) -- (-1.5, 0);
  %Oblique grid 
\pgftransformcm{1}{-1}{1}{1}{\pgfpoint{0cm}{-1.5cm}} \draw[style=help lines,dashed] (-3.3,-3.3) grid[step=1.5cm] (2,2); 
\end{tikzpicture}

\end{center}
\caption{\small \it The discrete model is defined on a rhombic lattice. We indicated the center $(\sigma,\tau)$ of the diamond $(\sigma\pm1,\tau\pm1)$. The equation of motion (\ref{ysystem}) expresses the variable at the top of the diamond in terms of the other three. }
\label{figure:LL}
\end{figure}
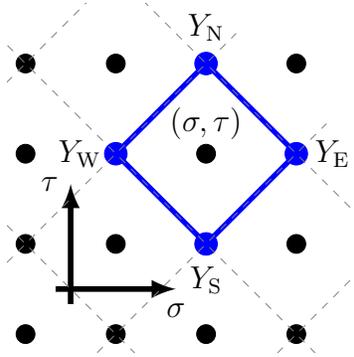

The time-evolution is discrete and specified as follows \cite{Faddeev}. We relabel the variables $\ff_n$ by means of two integers 
$\ff_{\sigma,\tau}$ with $\sigma+\tau =$ even, via $\ff_{2r,0} = \ff_{2r}$ and $\ff_{2r+1,1} = \ff_{2r+1}$. The relabeled variables specify the initial condition of the model. The time evolution will generate a discrete, cylindrical  1+1-D space time formed by a rhombic lattice. The time evolution proceeds via a local propagation rule \cite{Faddeev}.
We can focus on a single diamons shaped lattice cell
\beq
\label{news}
\fff_{\NN} \equiv \spc \ff_{\sigma,\tau+1}  \spc ,\qquad \fff_{\SS} \equiv \ff_{\sigma,\tau-1}\spc , \qquad
\fff_{\WW} \equiv\ff_{\sigma-1,\tau}\spc, \qquad \fff_{\EE}   \equiv \ff_{\sigma+1,\tau}.
\eeq
The evolution equation of the model reads 
\beq
\label{ysystem}
\spc \fff_{\NN}\spc 
\fff_{\SS}  \, = \,  \frac{\fff_{\WW} \fff_{\EE}}{(1 +  \fff_{\WW})(1+  \fff_{\EE})}  
\eeq
Equation (\ref{ysystem})  is the simplest example of a Y-system. It specifies the variable
$\fff_{\NN}$ at the top of the diamond shaped lattice cell in terms of the other three variables $\fff_{\EE},\fff_{\WW}$ and $\fff_{\SS}$, see fig. (\ref{figure:LL}).
The Y-system (\ref{ysystem}) defines an integrable lattice model, that can be recognized as  a discretized version of 2D hyperbolic geometry \cite{Faddeev}. The exchange relation (\ref{parafermion}) amounts to a quantization of this hyperbolic geometry.\footnote{\addtolength{\baselineskip}{-.5mm} In some way, one may view the model as a many body analogue of a hyperbolic billiard.}

The lattice model is a well defined quantum system, albeit one with a discrete time evolution. The model has been constructed \cite{Faddeev} so that in the large $L$ and IR limit, it describes a  2D continuum CFT with a non-linearly realized conformal symmetry with central charge $c = 1 + 6 (b + b^{-1})^2$ with $b^2 = {1}/{N}.$
As we will explain,  this CFT exhibits maximal Lyapunov behavior, and an infinite set of Ruelle resonances match the quasi-normal frequencies of the BTZ black hole \cite{BTZ-quasi-normal}.

It may seem surprising that an integrable model can display properties characteristic of many body quantum chaos.  To address this potential worry, one could choose to perturb the system away from integrability, e.g. by introducing frustration or by adding disorder.
 Since the features of quantum chaos will already become apparent in the unperturbed model, we will not go select among the list of such possible modifications\footnote{\addtolength{\baselineskip}{-1mm}One could add disorder e.g.  by using the freedom of normalization of the $\ff_n$ to set $
\ff_n{\!}^\dag\spc \ff_n\, =\, \kappa_n\smpc \mathbb{1}_{{}_{N\times N}},$ with $\kappa_{\rm n}$ random real numbers picked from a narrow probability distribution centered around  $\overline{\kappa_{\rm n}\!\!}\; = \kappa.$  Alternatively, one could add frustration e.g. by including a next-to-neighbor interaction in the time step rule (\ref{news}) and (\ref{ysystem}) via $\ff_{\sigma,\tau+1}\spc 
\ff_{\sigma,\tau-1}  \, = \, {(1 +  \epsilon\ff^{-1}_{\sigma+3,\tau})(1+  \epsilon\ff^{-1}_{\sigma-3,\tau})}/{(1 +  \ff^{-1}_{\sigma+1,\tau})(1+ \ff^{-1}_{\sigma-1,\tau})}.$} and instead focus on this idealized case, while ignoring the role of exact integrability. Indeed, we can note that there are other systems, such as  ${\cal N}=4$ SYM theory at large $N$, that are believed to be both integrable and chaotic. We will return to this point in the concluding section.

\medskip
\medskip

 {\it 2) Lyapunov from Goldstone} 
 
A central part of our reasoning consists of a new physical derivation of the Lyapunov behavior of an irrational CFT at finite temperature. The idea is as follows. 1+1-D CFTs are characterized by an infinite conformal symmetry group, given by reparametrizations 
of the lightcone coordinates $u$ and $v$
\beq
\label{ctrafo}
(u , v)\; \to \; (\xi(u),\eta(v))
\eeq
This conformal symmetry is broken by the conformal anomaly and by the presence of a finite energy density at finite temperature (and by the UV-cut-off).  For a CFT with a dense asymptotic energy spectrum, it is then natural to expect that the conformal symmetry is non-linearly realized in terms of a light Goldstone mode. 

\def\lL{\mbox{\tiny \sc l}}
\def\rR{\mbox{\tiny \sc r}}
This motivates us to consider the effective field theory of the relevant Goldstone excitation, described by the chiral field $\xi(u)$ in (\ref{ctrafo}) that parameterizes the conformal group. The effective Lagrangian is uniquely fixed by symmetries, and given by the geometric action of the Virasoro group \cite{alekseev}. 
 In section \ref{sec:Goldstone}, we will use this insight to derive the commutation relations of the Goldstone fields $\xi(u)$ and $\eta(v)$. We will find that the thermal expectation value of the commutators squared \beq
 \bigl\langle \,[\xi(u),\xi(0)]^2 \, \bigr\rangle \sim e^{2\lambda  u}, \qquad \qquad \bigl\langle \, [\eta(v),\eta(0)]^2\, \bigr\rangle \sim e^{2\lambda  v},
 \eeq
 initially grow exponentially with the time separation, with a temperature dependent Lyapunov exponent $\lambda = 2\pi/\beta$. In fact, we will derive the somewhat more precise result that, inside a thermal expectation value, the commutator between two generic local operators  takes the form\footnote{\addtolength{\baselineskip}{-1mm} Here for simplicity we only consider the time dependence of the correlator. In general, the left- and right-moving sectors each may have their own temperature and Lyapunov exponents $\lambda_{\lL,\rR} = {2\pi}/{\beta_{\lL,\rR}}.$}
\bea
\bigl[ \spc W(t_1 
), V(t_2 )\bigr]   \, \simeq \, \epsilon\spc e^{\lambda t_{12}} 
\partial_{t_1}\nspc
W(t_1)
 \spc \partial_{t_2}\nspc V(t_2) 
 \eea
with   $\epsilon$ some constant proportional to $1/c$. This result, which holds  for time-like separations 
in the intermediate range $c \gg \lambda t_{12}  \gg 1$, matches with the bulk interpretation of the commutator as resulting from a near horizon gravitational shockwave interaction \cite{shenker-stanford, kvv}.
\medskip

 {\it 3) Ruelle resonances as poles in OPE coefficients} 
 
A main characteristic of a chaotic system is that it thermalizes:
out of time ordered correlation functions decay to zero at late times. The approach toward equilibrium is governed by Ruelle resonances \cite{ruelle}. They appear as poles in the Fourier transform of the thermal two-point function, 
or in systems that obey the ETH \cite{ETH}, the matrix element  between two excited states with total energy $M$
\beq
\label{matrixelt}
G(\omega) = \int \! dt \, \langle M| \smpc \calo(t)\smpc \calo(0)\smpc |M\rangle\, e^{i\omega t}
\eeq 

The Ruelle resonances of holographic 2D CFTs are well studied \cite{BTZ-quasi-normal, kaplan}. As argued in \cite{fitzkap}, the matrix element reduces (for small $t$) to the thermal 2-point function. Its Fourier transform $G(\omega)$ has poles at resonant frequencies
\beq
\label{poles} 
\omega %_{{}_L} 
= - \frac{4\pi i} {\beta} (n + h),
\eeq
that coincide with the quasi-normal  modes of the BTZ black hole~\cite{BTZ-quasi-normal}. By factorizing the matrix element (\ref{matrixelt}) in the intermediate channel, we can write
\bea
G(\omega)\is \sum_{|i\rangle \in {\cal H}_{CFT}} \delta(M\nspc +\nspc \omega\nspc -\nspc E_i)
\, \bigl| \spc \langle M|\spc \calo\spc | \spc i\spc  \rangle \spc |^2  
\\[2.5mm]
\is  \rho(M\nspc +\nspc \omega) \,  |\spc \langle M|\spc \calo\spc \spc| \spc M\nspc + \nspc \omega \spc  \rangle\spc \bigr|^2 
\eea
where we used that in the Cardy regime, we can replace the spectral density  
$\rho(E) = \sum_{|i\rangle} \delta(E -\nspc E_i)$ by a continuous distribution, and label the CFT states by their energy. We learn that the Ruelle resonances dictate the analytic structure of the matrix element of a light operator  ${\cal O}$ between two highly excited states. 
This indicates that the resonances must show up as poles in  the OPE coefficient of a light operator and two heavy operators. Or in AdS-dual terms, the quasi-normal modes should show up as poles in the absorption and emission amplitudes of wave perturbations by a BTZ black hole. 

In section \ref{sec:Ruelle} we will show that the analytic continuation of the OPE coefficients of the continuum limit of our model indeed has poles located at the expected frequencies (\ref{poles}). This supports the statement that the continuum limit of the model is~ergodic.

\section{Lyapunov from Goldstone}\label{sec:Goldstone}

Consider an irrational 2D CFT with central charge $c\gg 1$ with an asymptotic density of states given by the Cardy formula, and with a sparse low energy spectrum. We place the CFT on a circle, parameterized by a periodic coordinate $x$ with period $2\pi$. We introduce light-cone coordinates $(u,v) = (t-x, t+x
 )$. 
 
 Consider a finite energy state with a constant expectation value for, say, the left-moving energy momentum tensor
\beq
\label{texp}
\bigl\langle T(u) \bigr\rangle \spc = \spc L_0 \; \gg\, \frac{c}{12}
\eeq
In this regime, we can associate to the state a finite inverse temperature $\frac{\beta}{2\pi} = \sqrt{\frac{c}{24 L_0}}.$

Let us perform a general conformal transformation (\ref{ctrafo}). We require that
\beq
\xi(u+2\pi) = \xi(u) + 2\pi
\eeq
The expectation value of the energy momentum tensor transforms non-trivially 
\beq
\label{ttrafo}
\bigl\langle T(u) \bigr\rangle \spc = \spc L_0\spc \spc \xi'{}^2(u) + \frac{c}{12}\spc S_\xi(u)
\eeq
with $S_\xi$ the Schwarzian derivative
\beq
S_\xi(u) =  \frac 1 2 \Bigl(\frac{\xi''(u)}{\xi'(u)}\Bigr)^2 - \Bigl(\frac{\xi''(u)}{\xi'(u)}\Bigr)' 
\eeq
The spontaneous breaking of conformal symmetry is displayed via the $\xi$-dependence of this expectation value. 
Indeed, we can compare the relation (\ref{ttrafo}) with the expression for the energy-momentum tensor of a fluid. The first term is analogous to the usual kinetic energy $\frac 12 \rho v^2$, whereas the second term in (\ref{ttrafo}) is the familiar vacuum contribution due to the conformal anomaly. It has a well-known physical explanation in terms of the Hawking-Unruh effect: the coordinate change from $u$ to $\xi(u)$ reshuffles the positive frequency (annihilation) and negative frequency (creation) modes, and thus alters the notion of the vacuum state.

Our physical assumption is that, for irrational CFTs at large $c$ and in the Cardy regime, it becomes accurate to treat the coordinate transformation $\xi(u)$ as a Goldstone field, in terms of which the conformal symmetry is non-linearly realized.
Adopting this logic, we thus promote $\xi(u)$ to an operator, that acts within the Hilbert subspace spanned by all states with energy density close to $L_0$, and their descendants.
Within this subspace, we can remove the expectation value in (\ref{ttrafo}) and elevate the equality in (\ref{ttrafo}) to an {\it operator identity}
\bea
\label{tident}
 T(u)  \is {L_0}\spc \xi'{}^2(u) + \frac{c}{12}\spc S_\xi(u).
\eea
As we will see shortly, the expression (\ref{tident}) for the energy-momentum tensor in terms of $\xi(u)$ is familiar from 
the geometric quantization of ${\rm Diff}(S^1)$, the group of (chiral) conformal transformations in 2D. 

A cautious reader may view equation (\ref{tident}) simply as a (in)convenient parameterization of the energy momentum tensor $T(u)$. Our assumption, however, is that
the symmetry parameter $\xi(u)$ acts as a genuine local quantum field that creates and annihilates local physical excitations. 
 Given that $\xi(u)$ is a scalar and $T(u)$ is the generator of conformal transformations, we know that\footnote{Here we absorb a factor of $\hbar\equiv {6/c}$ in the definition of $T(u)$. This is a customary step, that exhibits the fact that the commutation relations (\ref{txicom}) and (\ref{ttcom}) become semi-classical at large $c$.}  
\begin{eqnarray}
\label{txicom}
\bigl[T(u_1),\xi(u_2)\bigr] \is  \hbar\,  \xi'(u_2)\delta(u_{12})\qquad \qquad \hbar \equiv \frac 6 c \qquad\qquad\\[3.5mm]
\label{ttcom}
\bigl[T(u_1),T(u_2)\bigr]\is -\hbar(T(u_1)+T(u_2)) \delta'(u_{12}) + \frac \hbar {2} \delta{'''}(u_{12}).
\end{eqnarray}
The emergence of a light Goldstone mode at finite temperature can be explained as a physical consequence of the fact that an irrational CFT in the Cardy regime has an extremely dense energy spectrum. 

Equations (\ref{txicom}) and (\ref{ttcom}) become semi-classical in the large $c$ limit. From equation (\ref{texp}) we see that the field $\xi(u)$ has expectation value
\beq
\label{xiexp}
\bigl\langle \, \xi(u)\, \bigr\rangle \, = \, u
\eeq
So semi-classically, we can think of the Goldstone field as: $\xi(u) = u$ + small fluctuations. 

We are now ready to state the main technical result of this section:\\[1.5mm]
\noindent
{\it The three relations (\ref{tident}), (\ref{txicom}) and (\ref{ttcom})  uniquely determine the commutation relation of the Goldstone field $\xi(u)$, and are sufficient to derive the Lyapunov growth of commutators.}\\[2mm]
\noindent
Working to leading order in $1/c$, one finds that \cite{teschner-bootstrap-old}\cite{Dorn}
\smallskip
\bea
\label{dornone}
\bigl[ \xi(u_1), \xi(u_2)\bigr] \, = \,  \frac{\epsilon(u_{12})}{L_0} \, + \, \frac{\sinh(\lambda\spc \tau(u_1,u_2))}{L_0 \sinh \pi \lambda} \qquad \qquad \\[2.5mm]
\label{dorntwo}
\qquad \tau(u_1,u_2) \, =\,  \xi(u_1) - \xi(u_2) - \pi \epsilon(u_{12}), \qquad \quad
\lambda = \sqrt{{\frac{24 L_0}{c} }}
\eea
with $\epsilon(x)$ the stair step function, defined via $\epsilon'(x) = 2\delta(x)$ with $\delta(x)$ the periodic delta-function: $\epsilon(x)= 2n+1$ for $x \in (2\pi n, 2\pi(n+1))$. The same argument and derivation goes through for the right-movers. So we also have a right-moving Goldstone mode $\eta(v) = v$ + small fluctuations, that satisfies the analogous commutation~relation~(\ref{dornone}).\footnote{\addtolength{\baselineskip}{-1mm} For simplicity we will assume that the left and right movers have the same temperature.} The left- and right-moving Goldstone fields commute $[\xi(u),\eta(v)]=0.$

A detailed derivation of equation (\ref{dornone}) and (\ref{dorntwo}) can be found in \cite{teschner-bootstrap-old}\cite{Dorn}. Here we give a short summary. The constituent relation (\ref{tident}) between the energy-momentum tensor and the field $\xi(u)$ can be decomposed as
\bea
 \, T(u) \is \varphi'{}^2(u) -   2\varphi''(u), \\[2mm] 
\varphi(u) \is  \frac{\lambda} 2 \, \xi(u) + \frac 1 2  \log\bigl(\lambda \smpc \xi'(u)\bigr).
\label{phixi}
\eea
The commutation relations (\ref{txicom}) and (\ref{ttcom}) then follow from the free field commutator 
\beq
\label{phicom}
[\varphi(u_1),\varphi(u_2)] = \hbar \, \epsilon(u_{12}),
\eeq
with $ \hbar = 6/c$. So our task has been simplified: all we need to do is use relation (\ref{phixi}) to solve of $\xi(u)$ in terms of $\varphi(u)$, and use the chain rule to deduce the commutator of $\xi(u_1)$ and $\xi(u_2)$ from the free field commutator (\ref{phicom}) of $\varphi$. 

The free field $\varphi(u)$ is periodic up to a shift 
\beq 
\label{phiperiod}
\varphi(u+2\pi) = \varphi(u) + \pi \lambda. 
\eeq 
Using this fact,  equation (\ref{phixi}) integrates to \cite{Dorn}
\bea
\xi(u) \is \frac{1}{\lambda} \log \left(\spc \, \int_0^{2\pi}\!\!\!\! dy \, \frac{ e^{2\varphi(u+y)-\lambda\pi}}{\sinh\pi \lambda} \right) 
\eea
With this relation and equation (\ref{phicom}) in hand, it is now a relatively straightforward calculation to derive the result (\ref{dornone}) and (\ref{dorntwo}).

Let us turn to the physical consequences of equations (\ref{dornone}) and (\ref{dorntwo}).
We observe that $\lambda$ is equal to the maximal Lyapunov exponent $
\lambda ={2\pi}/{\beta}.$
We will assume that $\lambda\gg 1$, i.e. the thermal wave length is very short compared to the size of the spatial circle. The second term in the commutator (\ref{dornone}), and its right-mover counter part, thus grows exponentially with the coordinate differences $u_{12}$ and $v_{12}$ over the range 
\beq
\label{regime}
 \beta\;  \ll \; |u_{12}| \; < \; 2\pi, \qquad \qquad 
 \beta\;  \ll \; |v_{12}| \; < \; 2\pi.
\eeq 
We will restrict our attention to this coordinate range. In this regime, equation (\ref{dornone}) implies that the commutator between two local functions $\hat{f}(u_2) \equiv f(\xi(u_2))$ and $\hat{g} \equiv g(\xi(u_1))$ of the Goldstone fields satisfy 
\bea
\label{fgcom}
[\hat{f}(u_1), \hat{g}(u_2)] \!&\! \simeq\! & \!  e^{\lambda (|u_{12}|- 2\pi)} \; \hat{f}'(u_1)\spc \hat{g}'(u_2).
\eea
Here we used equation (\ref{xiexp}) to replace $\xi(u) \to u$ on the r.h.s.
 We would like to translate equation (\ref{fgcom}) into a statement about the commutator between local CFT operators.

Consider some local CFT operator ${\cal O}(u,v)$. Under the conformal transformation (\ref{ctrafo}) it transforms as
\bea
{\cal O}(u,v) & \rightarrow &  \xi'(u)^{h_{\lL}}\eta'(v)^{h_{\rR}}\spc  {\cal O}\bigl(\xi(u),\eta(v)\bigr)
\eea
Hence local operators are indeed non-trivial functions of the dynamical Goldstone fields. 

It is logical to take this observation one step further, and, similarly as we did for the energy-momentum tensor,
assume that local operators ${\cal O}(u,v)$ can be represented as c-number valued functions of the operator valued fields $\xi(u)$ and $\eta(v)$ and their derivatives.
The collection of these functions is determined by the spectrum and operator algebra of the CFT.  Their form is constrained by the locality requirement
 that space-like separated operators commute. This condition is very restrictive: it prescribes that primary local operators are all of the form \cite{Dorn, GN}
\bea
\label{localo}
{\cal O}_h(u,v)\!\!\!\!\!\!\!\!\!\!\!\!\!\!\! & & \; \;\; =\, \bigl(f(u,v)\bigr)^h,\\[4mm]
f(u,v) \is  \frac{\lambda^2\spc \xi'(u)\spc \eta'(v)}
{4\sinh^{2}\bigl(\frac \lambda 2 (\xi(u) - \eta(v))\bigr)}\, .
\label{fdef}
\eea
Equations (\ref{dornone}) and (\ref{dorntwo}) can then be used to compute the commutation relations between time-like separated operators, as follows.

The accepted test for Lyapunov growth of the commutator between two local operators $W$ and $V$ is to compute the expectation value 
\bea
\label{wvcom}
\bigl\langle W_\epsilon(u,v) \bigl[W(u,v), V(0,0)\bigr] V_\epsilon(0,0)\bigr\rangle
\eea
where the subscript ${}_\epsilon$ indicates a small displacement. This expectation value is equal to the difference between a time ordered and an out-of-time-ordered (OTO) correlation function. The OTO correlation function is obtained via analytic continuation of the time ordered correlation functions, where one circles, say, the coordinate $u$ around the origin. This operation amounts to analytic continuation of the left-moving conformal blocks to the second Riemann sheet. Of course, we could also choose to do the analytic continuation using the coordinate $v$. This would have given the same final result. 

The full-circle-monodromy ${\bf M}$ of a conformal block is the square ${\bf M}= {\bf R}^2$ of half-circle-monodromy known as the R-operation. The R-operator, acting on the left conformal blocks, re-orders the left-moving parts of the operators $W$ and $V$. In the linearized regime, i.e. to leading order in $1/c$, we can write ${\bf R} \simeq \mathbf{1} - {\bf r}$ with ${\bf r}$ the perturbative operation that takes the commutator between the left-moving parts of $W$ and $V$. 
The full-circle-monodromy is ${\bf M} \simeq {\bf R}^2= \mathbf{1}- 2 {\bf r}$ and thus the full commutator inside (\ref{wvcom}) is equal to acting with $(\mathbf{1}-{\bf M}) = 2{\bf r}$ on the two operators $W$ and $V$. From equation (\ref{fgcom}) we then deduce that
\beq
\label{wvcommutator}
\bigl[W(u_1,v_1), V(u_2,v_2)\nspc
\bigr]  
 \, \simeq \, 2\spc  e^{\lambda (u_{12}- u_0)}
 \; \partial_{u_1}\! W(u_1,v_1)\spc \partial_{u_2}\!V(u_2,v_2)
\eeq
This result, which holds for time like separation in the regime (\ref{regime}), displays the maximal Lyapunov behavior and the linearized 
gravitational effect of an early incoming perturbation (created by $V$)  on the arrival time of the outgoing signal (detected by $W$).

We end with a brief comment on the extension to higher orders. As indicated by the description of the monodromy moves, one expects that the commutator (\ref{wvcommutator}) exponentiates to a non-perturbative exchange relation. Fourier transforming the left-moving coordinate via $W_\alpha(v) = \int\! du\, e^{i\alpha u} W(u,v)$, this exchange algebra is expected to take the following form 
\beq
W_{\alpha}(v_1) V_{\omega-\alpha}(v_2) = \sum_{\beta}  M_{\alpha}{\!}^{\beta}\, V_\beta(v_2) \, W_{\omega-\beta}(v_1).
\eeq
If we assume that the bulk interaction is dominated by gravity, then AdS/CFT makes a precise prediction for the monodromy matrix $M_{\alpha}{}^\beta$ \cite{JMV}. The prediction precisely matches with the monodromy matrix of Liouville CFT \cite{JMV}. 

\def\Ez{\nspc\smpc\alpha\nspc\smpc}
\def\Eo{\nspc\smpc\epsilon_0\nspc\smpc}
\def\Et{\nspc\smpc\omega\nspc\smpc}
\def\Ef{\nspc\smpc\beta\nspc\smpc}

\section{A Chaotic Lattice Model}\label{sec:Lattice}

In this section, we will connect the FKV lattice model, defined by equations (\ref{parafermion}), (\ref{news}) and (\ref{ysystem}), with the above effective CFT derivation of Lyapunov behavior. 

The motivation for studying the lattice model is two-fold.  First, the geometric theory of the Goldstone fields $\xi(u)$ and $\eta(v)$ is an effective theory, that only becomes accurate  at finite temperature and long distance scales. Like all effective field theories, it does not define a fully consistent CFT by itself, nor does it have a unique UV completion. There are two ways in which one can try to embed an effective field theory into a self-consistent quantum system: a) look for an explicit UV completion, or b) introduce an explicit UV regulator. Approach b) is more practical. 

A second motivation is that one can hope that the lattice model, by virtue of being more well defined, may allow for more explicit dynamical understanding of the underlying mechanism for chaos. Indeed, it turns out that the lattice Liouville model can be formulated in a way that preserves the geometric appeal of the continuum theory \cite{Faddeev}

The Y-system (\ref{ysystem}) and the expression (\ref{localo}) of local operators in terms of the function (\ref{fdef}) both have a direct connection with hyperbolic geometry. To see this, we first note that the 1+1-D metric defined by
\bea
\label{ccmetric}
ds^2 = f(u,v) du dv = \frac{\lambda^2 d\xi d\eta}{\bigl(2\sinh(\frac{\lambda}{2}(\xi-\eta))\bigr)^2}
\eea
describes a hyperbolic space-time with constant negative curvature. The authors of \cite{Faddeev} gave a beautiful discretized description of this 2D hyperbolic metric as follows. 

We can write equation (\ref{ccmetric}) as
\bea
\label{cratio}
f(u,v) \is \frac 1 {\Delta^2}\, \frac{\bigl(e^{\lambda\xi(u+ \Delta)} - e^{\lambda\xi(u-\Delta)}\bigr)\bigl(e^{\lambda\eta(v+ \Delta)} - e^{\lambda\eta(v-\Delta)}\bigr)}{\bigl(e^{\lambda\xi(u+ \Delta)} - e^{\lambda\eta(v+\Delta)}\bigr)\bigl(e^{\lambda\xi(u-\Delta)} - e^{\lambda\eta(v-\Delta)}\bigr)}
\eea
with $\Delta$ an infinitesimal coordinate shift. Note that this expression for $f(u,v)$ looks like a cross-ratio. So it is invariant under M\"obius transformations. Now consider the values of $f(u,v)$ in four nearby points, separated by null shifts $\Delta$
\bea
\label{news2}
{\ff}_{\sigma,\tau-1} = f(u,v)\spc , \quad \ \ \ \ \; & & \  \ \ {\ff}_{\sigma+1,\tau}=f(u\nspc+\nspc\Delta,v)\spc ,\nonumber \\[-3.5mm]\\[-3.5mm]
{\ff}_{\sigma-1,\tau}= f(u,v\nspc+\nspc\Delta)\spc , \  \ & & \ \ \ {\ff}_{\sigma,\tau+1} = f(u\nspc+\nspc\Delta, v\nspc+\nspc\Delta) \spc . \nonumber
\eea
These four cross-ratios depend on six functions $e^{\lambda\xi(u)}, e^{\lambda\xi(u\pm \Delta)}$, $e^{\lambda\eta(v)}$, and $e^{\lambda\eta(v\pm \Delta)}$, but thanks to the M\"obius invariance, only three of the six functions are independent. Therefore, the four cross-rations (\ref{news}) satisfy one relation \cite{Faddeev}. Putting $\Delta=1$, it reads
\bea
\label{ytwo}
\spc \ff_{\sigma,\tau+1}\spc 
\ff_{\sigma,\tau-1}  \, = \,  \frac{\ff_{\sigma+1,\tau}\ff_{\sigma-1,\tau}}{(1 +  \ff_{\sigma+1,\tau})(1+  \ff_{\sigma-1,\tau})}.  
\eea
This confirms that the equation of motion of the FKV lattice model is a discretization of the hyperbolic metric (\ref{ccmetric}).
The parafermionic algebra 
\beq
\label{paratwo}
\ff_{\rm n} \ff_{\rm n\pm 1} \, =\, q^2 \spc \ff_{\rm n+1} \ff_{\rm n}
\eeq 
defines a quantization of the space of discretized hyperbolic metrics. 

Our new observation is that this lattice model can serve as a useful prototype of quantum chaos. The most direct way to substantiate this claim would be compute an out-of-time ordered four-point function of local operators
\bea
\label{otof}
\bigl\langle  \, \ff_{\sigma,\tau+t+1}\, \bigl[ \ff_{\sigma,\tau+t-1} ,  \ff_{\sigma,\tau+1} \bigr]\, \ff_{\sigma,\tau-1}\, \bigr\rangle_{\beta}
\eea 
at finite temperature, as a function of the time difference $t$.
While this would in principle be doable, we will leave this task to  future work. Instead we will cut the computation short, by banking on the results of \cite{Faddeev, Teschner-LL} that show that the above lattice model in the large $L$ limit approaches continuum Liouville CFT. Together with the result of the previous section, this is sufficient to demonstrate that the continuum limit of the lattice model displays maximal Lyapunov behavior.

\def\yy{{\mbox{\rm y\nspc\smpc}}}

For completeness, let us display a few more elements of the dictionary. Working to leading order at large $N$
\bea
\label{xone}
 e^{\varphi^+_{\rm n}} \spc e^{\varphi^+_{\rm m}} \spc =\spc 
 e^{\varphi^+_{\rm m}} \spc e^{\varphi^+_{\rm n}} \spc q^{2\epsilon_{\rm nm}}\ \; \ \ \, & &\ \ \qquad 
e^{\varphi(u_2)} e^{\varphi(u_1)} \spc =\, e^{\varphi(u_1)} e^{\varphi(u_2)} e^{\hbar \epsilon(u_{12})}  \, \nonumber \\[1.5mm]
e^{\varphi^+_{\rm n+L}}\spc =\spc e^{2\pi \lambda } \, e^{\varphi^+_{\rm n}} \qquad \ \ \ \; & &\ \   \qquad \quad\,\spc
e^{\varphi(u+ 4\pi)} \spc =\spc e^{2\pi \lambda} \spc e^{\varphi(u)} \\[1.5mm]
\textstyle \frac {\rm L} {2\pi} \spc e^{\varphi^+_{\rm n}}\, = \spc {e^{\frac  \lambda 2 \smpc \xi_{\rm n}}\! - e^{\frac \lambda 2\smpc \xi_{\rm n-1}}} \! \ \ \ \ & &\ \  \qquad \qquad\; \, e^{\varphi(u)}\, = \, \partial_u \spc e^{\frac \lambda 2\smpc \xi(u)}  \nonumber
\eea
The right column lists the formulas (\ref{phicom}), (\ref{phiperiod}) and (\ref{phixi}) that were used  to derive the commutation relation (\ref{dornone}) of the left-moving Goldstone variable $\xi(u)$. The left column is the lattice version of the same set of relations,
with $\epsilon_{\rm nm}$ the discretized stair-step function.
We can write a parallel set of formulas that represent the right-moving modes $\varphi(v)$ and $\eta(v)$ in terms of lattice variables $\varphi^-_{\rm n}$ and $\eta_{\rm n}$. 

Lattice variables $\varphi^\pm_{\rm n}$ that satisfy the exchange relation in (\ref{xone}) are obtained from the local operators $\ff_n$ in two steps \cite{Faddeev} \cite{Teschner-LL}. First we define two mutually commuting sets of chiral operators $\ww^\pm_{\rm n}$ via %\cite{Teschner-LL}
\beq
 \ww^+_{\rm n} \spc =\spc   \spc q\spc { \ff_{\rm 2n+1}\ff_{\rm 2n+2}^{-1}, }\qquad \qquad \ \ \ww^-_{\rm n}\spc = \spc q\spc \ff_{\rm 2n+1}\ff^{-1}_{2n}.
\eeq
These satisfy the algebra 
${\ww^\pm_{\rm n}\ww^\pm_{\rm m} = q^{\pm2\omega_{\rm mn}} \spc \ww^\pm_{\rm m}\ww^\pm_{\rm n}},$
with $\omega_{\rm mn} = {\rm sgn}(m-n)\delta_{|m-n|,1}$. The chiral variables $\varphi^\pm_{\rm n}$ are then defined as 
\bea
\varphi^\pm_{\rm n}  \is\rm \sum_m \epsilon_{nm} \log \ww^\pm_{m}, \qquad  \ 
\varphi^\pm_{\rm n+L}= \varphi^\pm_{\rm n} + 2\pi \lambda, \qquad
2\pi \lambda  =  \mbox{\large $\frac{1}{L}$} \sum_{n=1}^{L}\log \ww^\pm_n. 
\eea
At the initial time $\tau =0$, we can recover the single valued local parafermionic operators $\ff_{2\rm n}$ from the non-local chiral variables via
\beq
\label{factorize}
{\ff}_{\rm 2n}  \, = \, e^{\varphi^-_{\rm n}} e^{  \varphi^+_{\rm n}} 
\eeq
This is the lattice version of the relation $e^{2\phi(u,v)} = e^{\varphi(u) + \varphi(v)}$ that expresses a non-chiral free field vertex operator into the product of the two chiral vertex operators.  We note, however, that the time evolution (\ref{ytwo}) does not amount to free field propagation.

Among many other non-trivial results, \cite{Faddeev} and \cite{Teschner-LL} 
give an explicit construction of a unitary time evolution operator U that implements the time step (\ref{ytwo})
\bea
\ff_{\sigma, \tau+1} \, = \, {\rm U}^\dag\, \ff_{\rm \sigma,\tau-1} \spc {\rm U}.
\eea
This time evolution does not preserve the chiral factorization (\ref{factorize}).  However, it is shown that there exists a B\"acklund operator ${\rm B}$ that solves the time evolution via
\bea
\label{backlund}
{\ff}_{\sigma, \tau} \, =\,  {\rm B}^{-1} \spc e^{\varphi^+_{\! \raisebox{1pt}{\fontsize{1pt}{0.3pt}${\frac 1 2 }$}\! (\sigma-\tau)}} e^{ \varphi^-_{\raisebox{1pt}{\!\fontsize{1pt}{0.3pt}${\frac 1 2 }$}\!  (\sigma+\tau)}}\, {\rm B} 
\eea
This B\"acklund operation is causal but highly non-local, and no explicit representation of B is known at present. Indeed, as exemplified by this equation, all non-trivial dynamics of the Liouville lattice model is encoded in the way in which the two chiral sectors get mixed and become entangled under the time evolution step (\ref{ytwo}). Our results are evidence that this mixing and entangling is happening in a maximally efficient way.

Our argument that the lattice model exhibits maximal Lyapunov growth is a copy of the effective CFT derivation presented in section \ref{sec:Goldstone}.  The three relations in the left column of equation (\ref{xone}) specify the commutation relations of the $\xi_{\rm n}$ variables, in the same way as the right column fixes the commutator algebra of $\xi(u)$. The commutator algebra is expected to approach the continuum result (\ref{dornone}) in the large L limit. Our working assumption is that the exact solution (\ref{backlund}) of the lattice model leads to an expression of the local operators $\ff_{\sigma, \tau}$ in terms of the chiral modes $\xi_{\rm n}$ and $\eta_{\rm n}$ that mirrors formula (\ref{cratio}).
Via the same reasoning as in section \ref{sec:Goldstone}, this expression can then be used to verify that the lattice model is local and to establish that the OTO four point function (\ref{otof}) grows exponentially with time.

\section{Ruelle Resonances}\label{sec:Ruelle}
In this section we will expand on the topic of Ruelle resonances, which provide another signature of chaos and ergodicity. We will briefly review these concepts and then use the intuition for large $c$ irrational conformal field theories to translate the knowledge about these resonances into concrete CFT data. We will introduce a notion of 
OPE coefficients (of light operators between heavy states) as analytic functions of energy. We will see that the presence of Ruelle resonances, in combination with the conformal bootstrap and AdS/CFT, impose stringent constraints on the form of these analytic OPE functions. We will then verify that the known OPE coefficients of the effective CFT of section \ref{sec:Goldstone} and the continuum limit of the lattice model of section \ref{sec:Lattice} satisfy all these physical requirements.

\bigskip
\medskip

\noindent
{\bf 4.1 Ruelle resonances in CFT}

Ruelle resonances are poles in the Fourier transform of linear response functions that govern thermalization, the decay process towards  thermal equilibrium after a quench. Consider a small perturbation produced by a local operator $\calo_{\rm b}(x)$ to the Hamiltonian
\beq
\delta H = \int J(x) \calo_{\rm b}(x).
\eeq
Here $J(x)$ is an external source. Then one can study how this perturbation influences the time evolution of the expectation value of some other operator $\lb\calo_{\rm a}(0)\rb$, which for convenience we place at $x=0$. By expanding the evolution operator to linear order 
\beq
\label{response}
\delta \lb \calo_{\rm a}(0) \rb = \int\! dx'\, G^{\rm ret}_{\rm ab} (x') J(x'),
\eeq
where $G^{\rm ret}_{\rm ab}(x) = \theta(t) \langle [\calo_{\rm a}(x), \calo_{\rm b}(0)] \rangle$ (with $t$ = time component of $x$) is the retarded Green's function. $G^{\rm ret}_{\rm ab}(x)$  may be expressed in terms of two point functions as 
\beq
\label{gret}
G^{\rm ret}_{\rm ab}(x) = \theta(t) \bigl(G^+_{\rm ab}(x) - G^-_{\rm ab}(x)\bigr),
\eeq
with $G^+_{\rm ab}(x) = \langle \calo_{a}(x) \calo_{\rm b}(0)\rangle$ the time ordered two point function and $G^-_{\rm ab}(x) = \langle \calo_{b}(0) \calo_{\rm a}(x)\rangle$ the out-of-time-ordered two point function.
Equation (\ref{response}) is the basis of linear response theory, from which one can deduce transport properties such as the Kubo formula. Response functions are usually analyzed in the frequency domain. The Ruelle resonances appear as poles in the complex frequency plane. The imaginary part of the location of the poles determines the relaxation time. The leading behavior in $\delta \lb \calo_{\rm b} \rb (t)$ is governed the viscous hydrodynamical mode with the smallest imaginary part.

\begin{figure}[t]
\begin{center}
\begin{tikzpicture}[scale=1]
\path[draw, line width=1.5pt] (0.75, 0.5) -- (0.75, 3);
  \node[] at (0.75,3.25) {M};
   \node[] at (0.75,0.25) {M};
 \node[] at (0.35,1.55) {\Large{$\underset{h}{\sum}$}~~~};
 \path[draw, line width=1.5pt] (0.75,3.5*0.5) -- (1.5,3.5*0.5)
 node[pos=0.4,above]{$h$};
  \path[draw, line width=1.5pt] (1.5,3.5*0.5) -- (2.25,3.5*0.7)
 node[pos=1,right]{$a$};
 \path[draw, line width=1.5pt] (1.5,3.5*0.5) -- (2.25,3.5*0.3)
 node[pos=1,right]{$b$};
 %%%%
 \path[draw, line width=1.5pt] (5, 0.5) -- (5, 3)
 node[pos=0.4,left]{\Large{~=~~$\underset{\omega}{\sum}$}~~~}
  node[pos=0.5,right]{M\spc+\spc $\omega$};
   \node[] at (5,3.25) {M};
   \node[] at (5,0.25) {M};
 \path[draw, line width=1.5pt] (5,3.5*0.7)--(5+1.3,3.5*0.7)
 node[pos=1,right]{$a$};
  \path[draw, line width=1.5pt] (5,3.5*0.3)--(5+1.3,3.5*0.3)
 node[pos=1,right]{$b$};
 %%%%
 \path[draw, line width=1.5pt] (9.25, 0.5) -- (9.25, 3)
 node[pos=0.4,left]{\Large{~=~~$\underset{\omega'}{\sum}$}~~~}
  node[pos=0.5,right]{M\spc+\spc $\omega'$};
   \node[] at (9.25,3.25) {M};
   \node[] at (9.25,0.25) {M};
 \path[draw, line width=1.5pt] (9.25,3.5*0.7)--(9.25+1.3,3.5*0.7)
 node[pos=1,right]{$b$};
  \path[draw, line width=1.5pt] (9.25,3.5*0.3)--(9.25+1.3,3.5*0.3)
 node[pos=1,right]{$a$};
\end{tikzpicture}
\end{center}
\caption{\small \it Diagrammatical representation of crossing symmetry (given by the first identity) and the exchange algebra (given by the second) of the CFT correlation function of two heavy operators, labeled by $M$, and  two light ones, labeled by $a$ and $b$. }
\label{fig:crossing}
\end{figure}
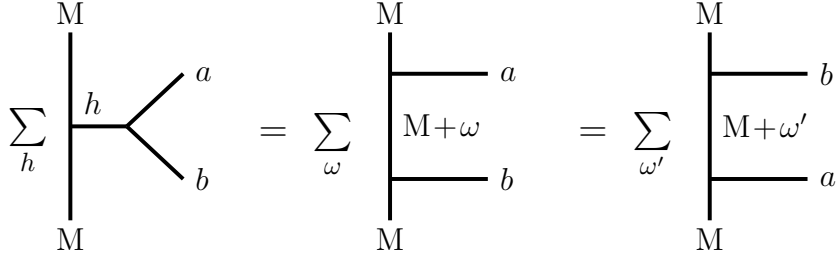

We are interested in studying this response function in a pure state microcanonical ensemble, defined by some highly excited CFT state $|M\rangle$ with a large scale dimension $M\gg~{\!\!\frac c{12}}$, so deep in the Cardy regime. The two-point functions of interest are given by the matrix~elements of the two light operators $\calo_{\rm a}$ and $\calo_{\rm b}$ between two heavy states 
\bea\label{Gdef}
G^+_{\rm a b}(u,v) \is \lb M | \calo_{\rm a}(u,v) \calo_{\rm b} (0) | M \rb \nonumber\\[-3.5mm]\\[-3.5mm]\nonumber
G^-_{\rm a b}(u,v) \is \lb M | \calo_{\rm b} (0) \calo_{\rm a}(u,v)  | M \rb
\eea
For 2D CFTs at large $c$, it has been argued in \cite{fitzkap} that the matrix elements (\ref{Gdef}) are dominated by the identity conformal block (which for $G^+(u,v)$ is given by the term with $h=0$ on the left in fig. \ref{fig:crossing}.) For large $c$, this identity block is well approximated by the thermal 2-point function on an infinite 1D space
\bea
\label{gthermal}
G^\pm_{\rm ab}(u,v) \! & \! \simeq \! & \! \delta_{\rm ab}\left(\frac{\pi/\beta}{\sinh\nspc\bigl(\frac{\pi}{\beta} (u\nspc \pm \nspc i\epsilon) \bigr)}\right)^{\! 2h} \left(\frac{\pi/\beta}{\sinh\nspc\bigl(\frac{\pi}{\beta} (v\nspc\pm \nspc i\epsilon) \bigr)}\right)^{\! 2h}.
\eea 
with $\beta = \pi \sqrt{c/6M}$. This is a useful result, that supports both the ETH and the dual identification of the two point function as the boundary-to-boundary propagator of a bulk field in a BTZ black hole background. 

The validity of equations (\ref{gthermal}) is somewhat limited, however. It only holds for spatial separations that are small compared to the size of the spatial circle, and for the OTO two-point function, the time difference must be short compared to the scrambling time, since otherwise one enters the Lyapunov regime. On the gravity side, the perturbation $\calo_{\rm b}$  creates an incoming wave that  may collide with the outgoing wave detected by $\calo_{\rm a}$, and thereby substantially affect its future trajectory.
This gravitational effect will show up as a modification of the OTO two-point function $G^-(u,v)$, and was studied in section \ref{sec:Goldstone}. Here we will focus on the late time behavior of the time ordered 2-point function $G^+(u,v)$.

The incoming wave deforms the black hole horizon state. The subsequent ring down of the black hole towards equilibrium is the dual of the thermalization process of the CFT. Both processes are governed by an infinite set of resonances. On the gravity side, these resonances are the quasi-normal modes. These can be analyzed perturbatively, by considering small fluctuations of fields propagating in the neighborhood of the black hole horizon. This resonant quasi-normal frequencies are an infinite series of complex numbers, labeled by a non-negative integer $n$ via \cite{BTZ-quasi-normal}
\beq\label{qnf}
\omega =\pm k - i\frac{4\pi }{\beta}(n+h).
\eeq
with $k$ is the momentum of the infalling mode and $h$ the conformal dimension of the fluctuating field. This result was derived using the Poincar\'e patch, corresponding with a CFT on an infinite line, and with vanishing Dirichlet boundary conditions at infinity \cite{BTZ-quasi-normal}.\footnote{\addtolength{\baselineskip}{-1mm} Notice that this Dirichlet boundary condition eliminates all gravitational excitations corresponding to the Virasoro descendants in the CFT. This restriction will become relevant later.}
It is reasonable to assume that the result generalizes to black holes in global AdS, with a periodic spatial boundary, by replacing the momentum $k$ by an integer angular momentum~$\ell$.

In the CFT, the quasi-normal modes manifest themselves as Ruelle resonances, that appear as poles in the Fourier transform of the retarded thermal Green's function (\ref{gret})
\bea
\label{gfourier}
G^{\rm ret}_{\rm ab}(\omega,\ell) \is \int\!\! du\! \int\!\! dv \; e^{i\frac 1 2 (\omega + \ell )u} e^{i\frac 1 2 (\omega - \ell )v}\; G^{\rm ret}_{\rm ab}(u,v),
\eea
which via equation (\ref{gthermal}) yields a spectrum that matches with the gravity prediction (\ref{qnf}).
Our goal in this section is to use the presence of these Ruelle poles to extract useful information about the OPE coefficients of the CFT. Earlier paper with results that overlap with this section are \cite{fitzkap, Solodukhin}.

\bigskip
\medskip

\noindent
{\bf 4.2 Resonances and OPE coefficients}

As  a preparation, let us look at the different conformal block expansions of the matrix elements (\ref{Gdef}), as shown schematically in fig. (\ref{fig:crossing}). 
The first equal sign of these identities represents the crossing symmetry relation\footnote{Here we temporarily rotate to euclidean signature and set $(u,v) = (z,\overline{z})$.}
\bea
\label{bpzbootstrap}
G_{\rm ab}(z) \, = \, \sum_{\rm \omega} \;\CC
_{\rm a M M\plus\omega} \CC^{\rm {M\plus\omega}}_{\rm \,M b}\; \bigr| \spc {\cal F}_{{\!\nspc}_{\rm M\plus\omega}}  \bigl[\!\! \begin{array}{cc} \mbox{\scriptsize M} \!\!&\!\! \mbox{\scriptsize M}\\[-3.5mm] \raisebox{1pt}{\scriptsize a} \!\! &\!\! \raisebox{1pt}{\scriptsize b} \end{array} \!\!\bigr]\spc 
(z) \bigr|^2 \is  \sum_{\rm h} \;\CC_{\rm MM}^{\, \rm h}\, \CC_{\rm {a\smpc h \smpc b}}\; \bigr| \spc {\cal F}_{\raisebox{-2pt}{\scriptsize\nspc ${\rm h}$}}  \bigl[\!\! \begin{array}{cc} \mbox{\scriptsize M} \!\!&\!\! \mbox{\scriptsize a}\\[-3.5mm] \raisebox{1pt}{\scriptsize M} \!\! &\!\! \raisebox{1pt}{\scriptsize b} \end{array} \!\!\bigr](1-z) \spc \bigr|^2\spc\nonumber\\[-3mm]
 %\Bigr|\spc \psi\bigl(\tbac\bigr)(w) \cdot \psi\bigl(\abc\bigr)(z)\spc 
\eea
where ${\cal F}_h  \bigl[\!\! \begin{array}{cc} \mbox{\scriptsize M} \!\!&\!\! \mbox{\scriptsize M}\\[-3.5mm] \raisebox{1pt}{\scriptsize a} \!\! &\!\! \raisebox{1pt}{\scriptsize b} \end{array} \!\!\bigr]\spc 
(z) $ represents the Virasoro block shown on the left in fig. (\ref{fig:crossing}). We see that crossing symmetry relates the `t-channel block' with  heavy intermediate channel (labeled by M+
$\omega$)  to the `s-channel block' with a  light intermediate channel (labeled by $h$). 

The second relation in fig. (\ref{fig:crossing}) is the exchange algebra relation,
\bea
\label{bpzbootstrap2}
\sum_{\rm \omega} \spc \CC_{\rm a\smpc M\smpc M\mbox{\tiny+}\omega}\spc \CC_{\rm\spc Mb}^{\rm M\plus\omega}\bigr| \spc {\cal F}_{{\!\!\nspc}_{\rm M\plus\omega}} \! \bigl[\!\! \begin{array}{cc} \mbox{\scriptsize M} \!\!&\!\! \mbox{\scriptsize M}\\[-3.5mm] \raisebox{1pt}{\scriptsize a} \!\! &\!\! \raisebox{1pt}{\scriptsize b} \end{array} \!\!\bigr]\spc 
(z) \bigr|^2 \is \sum_{\rm \omega'} \CC_{\rm {b\smpc M\smpc M\plus\omega'}}\spc \CC_{\rm\, Ma}^{M\mbox{\tiny +}\omega'}\spc \bigr| \spc {\cal F}_{{\!\!\nspc}_{\rm M\plus\omega'}} \!\nspc \bigl[\!\! \begin{array}{cc} \mbox{\scriptsize M} \!\!&\!\! \mbox{\scriptsize M}\\[-3.5mm] \raisebox{1pt}{\scriptsize b} \!\! &\!\! \raisebox{1pt}{\scriptsize a} \end{array} \!\!\bigr]\spc
(1/z) \spc \bigr|^2, %\Bigr|\spc \psi\bigl(\tbac\bigr)(w) \cdot \psi\bigl(\abc\bigr)(z)\spc 
\eea
 that imposes locality in the Euclidean region. In Lorentzian language, it implies that the R-matrix $R_{\omega,\omega'}$ that relates the chiral time-ordered conformal block (labeled by M\spc+\spc$\omega$) to the out-of-time-ordered conformal block (labeled by M\spc+\spc $\omega'$) is an appropriate unitary
transformation, so that in the euclidean region, it cancels out between the left- and right-movers of the complete CFT four-point function. After rotating to Lorentz signature, the R-matrix does show up in a non-trivial way, in the relation between the time-ordered Green's function $G^+_{\rm ab}(u,v)$ and the OTO Green's function $G_{\rm ab}^-(u,v)$ \cite{JMV}.

We wish to extract information regarding the Fourier transform of  $G_{\rm ab}$ from its expansion (\ref{bpzbootstrap}) in conformal blocks, in the channel shown in the middle of fig.~(\ref{fig:crossing}). This is not directly possible, since no explicit expression for the Virasoro conformal blocks is known. So let us take a step back and write the crossing symmetry formula as a sum over primary operators and descendants. Let $\gamma^\omega_{n}$  denote the $n$-th coefficient of the Laurent expansion of the Virasoro conformal block   ${\cal F}_{{\!\!\nspc}_{\rm M\plus\omega}} \! \bigl[\!\! \begin{array}{cc} \mbox{\scriptsize M} \!\!&\!\! \mbox{\scriptsize M}\\[-3.8mm] \raisebox{1pt}{\scriptsize a} \!\! &\!\! \raisebox{1pt}{\scriptsize b} \end{array} \!\!\bigr](z)$. From now on we focus on the diagonal part of the two point function $G_{\rm ab}(u,v) = \spc G(u,v) \spc \delta_{\rm ab}$. It has the following expansion 
\beq
G(u,v)\, =\,   \sum_{|i\rangle} G_{\omega_{i,L}}(u) \spc G_{\omega_{i,R}}(v) 
\eeq
\beq
G_{\omega_L}(u)\, =\, \sum_{n_L} \CC_{\rm a \smpc M \smpc M\plus\omega_{L}}~ \gamma^{\omega_L}_{n_L}~e^{i(\omega_{L} +n_L)u} 
\eeq 
and a similar formula holds for $G_{\omega_R}(v)$. Here $|\spc i\spc \rangle$ runs over all conformal primary states of the CFT in the neighborhood of the high energy state $|M\rangle$. In the sum we allowed all states with different left- and right conformal dimension $(\Delta_{i,L},\Delta_{i,R}) = (M+\omega_{i,L}, M+\omega_{i,R})$.

We want to take the Fourier transform  (\ref{gfourier}) with respect to both light-cone coordinates.
It is useful to introduce the spectral density of CFT primary states 
\beq
\rho(\omega_L,\omega_R) = \sum_{|i\rangle} \delta({\rm M}+\omega_L- \Delta_{i,L}) \delta({\rm M}+\omega_R- \Delta_{i,R}). 
\eeq
We then have
\beq\label{gretdef}
G^{\rm ret}(\omega,\ell) = \int\! \frac{d\omega'}{2\pi} \, \frac{G(\omega',\ell)}{\omega'-\omega-i\epsilon}
\eeq
\bea
\label{goal}
G(\omega,\ell) \is \sum_{n_L,n_R} \;\, \CC_{{\rm a\smpc M \smpc M}\plus\omega_L-n_L} \CC_{{\rm a\smpc M \smpc M}\plus\omega_R-n_R}~\gamma^{\omega_L}_{n_L}~\gamma^{\omega_R}_{n_R}\; \rho(\omega_L \nspc - \nspc n_L,\omega_R-n_R),
\eea
with $\omega_L = \frac 1 2 (\omega+\ell)$ and $\omega_R =  \frac 12 (\omega-\ell)$.
For a given CFT,  $G^{\rm ret}(\omega,\ell)$ contains exact information about the spectrum of primary fields, in the form of a dense set of poles along the real axis, with residues equal to the corresponding OPE coefficient. The Ruelle resonances appear as a series of poles in  $G^{\rm ret}(\omega,\ell)$ located off the real axis. Based on equation (\ref{gthermal}) and the results of \cite{fitzkap} and \cite{BTZ-quasi-normal}, we expect that their location should match with the quasi-normal frequencies (\ref{qnf}). 

The spectrum of an irrational CFT at large $c$ becomes very dense in the Cardy regime. In this type of situation, it is customary to treat the spectrum as a continuum with spectral density given by the Cardy formula, and elevate the OPE coefficients to continuous functions of the conformal weights. The Ruelle resonances are then expected to arise as poles in the analytic continuation of the OPE coefficients.\footnote{\addtolength{\baselineskip}{-.5mm} Evidently, the Ruelle poles do not arise from the density of states. The Laurent coefficients $\gamma^\omega_{n}$ are fixed by conformal symmetry and they only exhibit poles for frequencies associated to degenerate states. The degenerate states appear at different locations than the quasi-normal modes.}

\smallskip

Let us summarize. The OPE coefficients between light and heavy operators satisfy several non-trivial compatibility conditions:
they solve the CFT bootstrap equations (\ref{bpzbootstrap}) and (\ref{bpzbootstrap2}), and must be compatible with the known location (\ref{qnf})
of the Ruelle resonances. The question is: do these conditions uniquely fix the form of the OPE coefficients, in the universal high energy regime in which the CFT spectrum is governed by the Cardy formula? Do we know of any solutions to these conditions?

\smallskip

\bigskip
\medskip

\noindent
{\bf 4.3 Ruelle from Liouville}

The answer to the last question is affirmative:  Liouville theory solves both conditions. The bootstrap program of Liouville CFT is by now on firm footing \cite{teschner-bootstrap}. Our new observation is that the OPE coefficients of Liouville CFT, given by the famous DOZZ formula \cite{DOZZ}, indeed exhibit a series of poles that precisely match with the quasi-normal frequencies (\ref{qnf}) of the BTZ black hole. This observation gives extra support to the proposal that Liouville theory should be viewed as the effective CFT that captures universal high energy behavior of holographic CFTs. As we will discuss in the concluding section, this result also sheds light on whether the lattice model of section \ref{sec:Lattice} has ergodic dynamics or not.

Liouville CFT has a continuous spectrum labeled by the momentum variable $\alpha$ via $\Delta_\alpha = \alpha(Q-\alpha)$ with $c = 1+ 6Q^2$ and $Q=b+b^{-1}$.
In Appendix \ref{DOZZ} we review the expression for the three point function $C(\alpha_1,\alpha_2,\alpha_3)$ for a light operator,  labeled by $\alpha_1$, and two heavy operators, labeled by $\alpha_2$ and $\alpha_3$. Denoting the conformal dimensions as $\Delta_{1}=h$, $\Delta_2 $ = M and $\Delta_3 =$ M\spc +\spc $\omega$, the corresponding  Liouville momenta are
\bea
\label{alpharels}
\alpha_1\, \simeq \, bh, \qquad & &  \alpha_{3} - \alpha_2 \, \simeq \,i  \frac{\omega}{2\sqrt{{\rm M}}}\, = \, i b\, \frac{\beta}{4\pi}\, \omega % \qquad \qquad \lambda =  \frac{2\pi}{\beta}
\eea
where $\beta = 2\pi/b\sqrt{{\rm M}}$ is the inverse temperature associated with the state $M$.

The DOZZ three-point function $C(\alpha_1,\alpha_2,\alpha_3)$ has a rich pole structure. As explained in Appendix \ref{DOZZ}, the series of poles that are relevant to our physical situation are located at
\beq
\label{dozzpoles}
\alpha_1 + \alpha_3-\alpha_2 = n b, \qquad \ n\in \mathbb{Z},
\eeq  
which via equations (\ref{alpharels}) and (\ref{dozzpoles}) tells us that $C_{\rm h \spc M \spc M+\omega}$ has poles at  
\bea
\omega \is -i \frac{4\pi}{\beta}(n+h). %$~~~~n\in\mathbf{Z}.
\eea
Plugging this into (\ref{goal}), and doing the integral (\ref{gretdef}), we learn that the retarded Green's function $G^{\rm ret}(\omega,\ell)$ has poles for 
\bea
\label{ruelle-liouville}
\qquad \qquad \qquad \omega \is  -\ell + 2n_{L} - i \frac{4\pi}{\beta}(n+ h),\nonumber \\[-3mm] & & \qquad \qquad \qquad \qquad \qquad \qquad \qquad n \in \mathbb{N}. \\[-3mm]
\qquad \qquad \qquad \omega \is \ell + 2n_{R} - i \frac{4\pi}{\beta}(n+ h) .\nonumber
\eea
These are the Ruelle resonances that govern the thermalization dynamics of Liouville CFT. Notice that relative to the list (\ref{qnf}) of BTZ quasi-normal modes, 
the series (\ref{ruelle-liouville}) reveals additional poles shifted by the excitation numbers $n_L$ and $n_R$ of the left- and right-moving Virasoro descendants.
These additional poles arise because in our CFT calculation, we did not exclude the possibility that the incoming wave created by $\calo_{\rm a}$ also excites boundary gravitons. If we ignore the  energy stored in the boundary gravitons, we recover the expected BTZ result (\ref{qnf}).

\section{Conclusions}\label{sec:Conclusion}

In this note we have made three observations that clarify the geometric origin of chaotic behavior in irrational 2D CFTs. We argued that in holographic CFTs at finite temperature, conformal symmetry is non-linearly realized by means of universal Goldstone-like fields $\xi(u)$ and $\eta(v)$, that describe the near-horizon gravitational dynamics of the dual theory.
The effective field theory is weakly coupled and its maximal Lyapunov behavior can be demonstrated at the semi-classical level. 

We used this insight to propose a new toy model for quantum chaos in the form of the FKV lattice model,  with an integrable equation of motion given by a Y-system.
Integrability may seem unhelpful for generating ergodic behavior. Indeed, integrable systems are seen as prototypical counter-examples for the ETH:
their single state microcanonical ensemble is understood to be described by the generalized Gibbs ensemble (GGE), which has many chemical potentials, one for each conserved quantity \cite{GGE}. 
However, this reasoning assumes that the state that defines the microcanonical ensemble is an (approximate) eigenstate of many or all conserved quantities. Instead, if we choose an energy eigenstate that otherwise is a random linear superposition of eigenstates of all other conserved quantities, then the usual ETH can still apply. The conserved quantities in 
the FKV lattice model are highly non-local, and with respect to local observables the dynamics still looks random and thermalizing. As discussed in the introduction, this random dynamics can be reinforced by introducing some degree of disorder.

Indeed the discrete model seems particularly useful for studying propagation of entanglement, and even though the continuum limit is expected to described by a CFT, the entanglement propagation generated by the Y-system rule (\ref{ysystem}) is non-ballistic and mixes left- and right-moving signals. Our conjecture that the lattice dynamics is ergodinc is further supported by the fact that the continuum limit of the model is expected to be described by Liouville theory, which via the observation of section \ref{sec:Ruelle} has Ruelle resonances that 
prescribe the approach towards thermal equilibrium.

Of course, underlying all three observations in this note, is the idea that the bulk gravitational dynamics of holographic  2D CFTs is accurately captured by 2D Liouville CFT \cite{JMV}.
This emergent Liouville field can be viewed as encoding the dynamical interplay between geometric entanglement and energy flow. This interpretation combines the idea of kinematic space \cite{kinematic}, that the entanglement entropy $S(u,v)$ of an interval $[u,v]$ between two space-like separated points $x=u$ and $x=v$ describes a metric on a 2D hyperbolic space via 
\beq
ds^2 = \partial_u \partial_v S(u,v)\spc du dv,
\eeq
with the first law of entanglement thermodynamics 
\beq
\label{thermo-relation}
\delta S(u,v) = \delta K(u,v) = \int_u^v\! dx\, P_{[u,v]}(x)\spc \delta T_{00}(x)
\eeq
with $P_{[u,v]}(x)$ the conformal Killing vector associated with the Rindler Hamiltonian $K(u,v)$ of the interval $[u,v]$.
Equation (\ref{thermo-relation}) can be integrated \cite{nele} into an expression for the energy-momentum tensor $T_{\alpha\beta}$ in terms of the entanglement entropy $S(u,v)$,
which looks exactly like the Liouville energy momentum tensor, via the identification
\bea
\phi(u,v) = S(u,v)
\eea
 of the Liouville field with the entanglement~entropy. Note that both quantities define locally constant curvature metrics, and both transform inhomogeneously under coordinate transformations. Hence the dynamics of kinematic space seems intimately connected with the emergence of an effective Liouville field in holographic 2D CFT.

\medskip

\begin{center}
{\bf Acknowledgements}
\end{center}
\vspace{-2mm}

We thank Nele Callebaut, Bruno le Floch, Aitor Lewkowycz, Lauren McGough, Mark Mezei, Juan Maldacena, Eric Perlmutter, Eliezer Rabinovici and Douglas Stanford for helpful discussions.
This research of H.V. is supported by NSF grant PHY-1314198.

\begin{appendix}
\section{DOZZ three point function}\label{DOZZ}
In this Appendix we summarize the DOZZ formula for the OPE coefficients of Liouville theory \cite{DOZZ}. A nice review can be found in \cite{harlow-witten}. After introducing the formula we will study its analytic properties which are relevant for the application we consider in the main text. 

The DOZZ formula computes the OPE coefficients between three primary operators of Liouville theory. These operators are labeled by a complex parameter $\alpha$ and can be written in terms of the Liouville field $\varphi(z,\zb)$ in the following way
\beq
V_{\alpha_j}(z,\zb) =e^{ 2 \alpha_j \varphi(z,\zb)},~~~~j=1,2,3.
\eeq
The dimension of the state in term of its label is $\Delta_\alpha = \overline{\Delta}_\alpha = \alpha (Q-\alpha)$. As usual the central charge is $c=1+6 Q^2$ and $Q=b+\frac{1}{b}$, where $b$ is a positive real parameter. The semiclassical limit $c\gg 1$ corresponds to $b\ll 1$. That is the limit we are interested in, although the result for Liouville theory is supposed to be valid more generally. Now we can state the DOZZ formula which for generic $\alpha_{1,2,3}$ and $b$ is given by 
\beq\label{DOZZf}
C(\alpha_1,\alpha_2,\alpha_3) =  \frac{\left[ \pi \mu \gamma(b^2) b^{2-2b^2}\right]^{\left(Q-\sum_i \alpha_i\right)/b}\Upsilon_0 \Upsilon_b(2\alpha_1)\Upsilon_b(2\alpha_2)\Upsilon_b(2\alpha_3)}{\Upsilon_b(\sum_i \alpha_i-Q)\Upsilon_b(\alpha_1+\alpha_2-\alpha_3)\Upsilon_b(\alpha_1-\alpha_3-\alpha_2)\Upsilon_b(\alpha_2+\alpha_3-\alpha_1)},
\eeq
where $\mu$ is the cosmological constant, $\gamma(x)\equiv \Gamma(x)/\Gamma(1-x)$ and $\Upsilon_0 = \frac{d \Upsilon_b(x)}{dx}\big|_0$. $\Upsilon_b(x)$ is an entire function. It is usually defined by analytic continuation of an integral representation valid for $0<{\rm Re}(x)<Q$ which can be found in \cite{DOZZ}. Here we will not need more information about this function other that its zeros since it gives the position of the poles in the DOZZ formula in terms of the $\alpha$'s. Specifically, they are located at 
\beq
x=-\frac{m}{b} - b n, ~~~x=\frac{m'+1}{b} +b (n'+1),~~~m,m',n,n'\in \mathbb{Z}^+.
\eeq
Looking at the formula (\ref{DOZZf}) we see all the poles are located in terms of the labels $\alpha$ at 
\bea
\alpha_1+\alpha_2+\alpha_3 -Q&=& - \frac{m}{b} - b n,~~~{\rm or}~~~\frac{m'+1}{b} + (b+1) n',\\
\alpha_1+\alpha_2-\alpha_3 &=&- \frac{m}{b} - b n,~~~{\rm or}~~~\frac{m'+1}{b} + (b+1) n',\label{polo1}\\
\alpha_1-\alpha_2+\alpha_3 &=&- \frac{m}{b} - b n,~~~{\rm or}~~~\frac{m'+1}{b} + (b+1) n',\label{polo2}\\
\alpha_2+\alpha_3-\alpha_1 &=&- \frac{m}{b} - b n,~~~{\rm or}~~~\frac{m'+1}{b} + (b+1) n'.
\eea

These are all the poles of the OPE coefficients. Now we will use the semiclassical limit to identify the poles that are physically relevant for the discussion in the main text, i.e. the ones that survive the $b\to 0$ limit. The external operators that we are interested in are such that one is light, $\alpha_1$, two are heavy, $\alpha_2$ and $\alpha_3$, and the difference between the two heavy operators is small. This means we fix the scaling with $b$ in the $b\to 0$ limit such that $\alpha_1 \sim b$, $~\alpha_2\sim \alpha_3\sim b^{-1}$ and $\alpha_3-\alpha_2 \sim b$. Then it is clear that the relevant poles to retain are the ones in equation (\ref{polo1}) and (\ref{polo2}) for only $n$ non zero. These two sets of pole for $n>0$ can be combined into a single formula 
\beq
\alpha_1+\alpha_3 - \alpha_2 =  b n,~~~~n\in\mathbb{Z},
\eeq
where now $n$ runs over all the integers. We see this has the right scaling since both the left and right hand side scale as $b$ in the semiclassical limit. All the rest of the poles disappear in the heavy-heavy-light limit we are interested in here. 

\end{appendix}
 \begingroup\raggedright\endgroup

%\bibliography{trapbib}
\end{document}